\documentclass[sn-mathphys]{sn-jnl}
\catcode`\|=12\relax

\algrenewcommand\alglinenumber[1]{\footnotesize #1}
\algnewcommand\AAnd{\textbf{and} }
\algnewcommand\AOr{\textbf{or} }
\algnewcommand\AIN{\textbf{in} }
\algnewcommand\ATO{\textbf{to} }

\usepackage{booktabs,enumitem,graphicx,multirow,siunitx,subcaption,textcomp}

\usepackage{tikz,xfp}
\usepackage[outline]{contour}
\usetikzlibrary{arrows,calc,matrix,positioning,shadows,shapes.geometric}

\begin{document}

\title[Fiuncho]{
  Fiuncho: a program for any-order epistasis detection in CPU clusters
}

\author*[1]{\fnm{Christian} \sur{Ponte-Fernández}}\email{christian.ponte@udc.es}
\author[1]{\fnm{Jorge} \sur{González-Domínguez}}
\author[1]{\fnm{María J.} \sur{Martín}}

\affil[1]{
  \orgname{Universidade da Coruña},
  \orgdiv{CITIC, Computer Architecture Group},
  \orgaddress{
    \street{Facultad de Informática, Campus de Elviña},
    \city{A Coruña},
    \postcode{15071},
    \country{Spain}%
  }%
}

% Abstract: 100 to 150 words
\abstract{
  Epistasis can be defined as the statistical interaction of genes during the
  expression of a phenotype. It is believed that it plays a fundamental role in
  gene expression, as individual genetic variants have reported a very small
  increase in disease risk in previous Genome-Wide Association Studies. The most
  successful approach to epistasis detection is the exhaustive method, although
  its exponential time complexity requires a highly parallel implementation in
  order to be used. This work presents Fiuncho, a program that exploits all
  levels of parallelism present in \textit{x86\_64} CPU clusters in order to
  mitigate the complexity of this approach. It supports epistasis interactions
  of any order, and when compared with other exhaustive methods, it is on
  average 358, 7 and 3 times faster than MDR, MPI3SNP and BitEpi, respectively.
}

% Keywords: 4 to 6

\maketitle

\section{Introduction}

With the proliferation of next-generation sequencing technologies, the cost of
sequencing genomes has been reduced, and Genome-Wide Association Studies (GWAS)
have become more popular. GWAS are observational studies that attempt to
decipher the relationship between a particular trait or phenotype and a group of
genetic variants from several individuals. Much of the early work in GWAS
considered genetic variants in isolation, and the results of those studies were
unsatisfactory for the task at hand. The studies commonly reported associations
with variants of unknown significance that increased disease risk at very low
levels, and thus their usefulness in clinical applications was
limited~\cite{genin_2020_missing}. One hypothesis that explains this outcome is
a phenomenon called epistasis: the statistical interaction of genes among
themselves, or with the environment, during the expression of a phenotype so
that individual variants by themselves display little to no association with
said phenotype. Nevertheless, looking for epistatic interactions instead of
individually associated genetic markers is a much more complex task, and it is
still an actively researched field.

A multitude of methods for detecting epistasis have been proposed in the
literature. In essence, these methods seek to identify the combination(s) of
variants that best explain the phenotype outcome observed in the data. This is a
computationally intensive problem with a complexity that scales exponentially
with the number of variants in combination considered (also known as the
epistasis order) and the number of variants included in the input data. As a
consequence of that, the methods developed followed two different approaches:

\begin{itemize}
  \item Exhaustive methods: all genetic variant combinations from the input data
  (up to a certain size or interaction order) are tested for epistasis.

  \item Non-exhaustive methods: a fraction of the genetic variant combinations
  are tested, following a particular heuristic that reduces the search space.
  Non-exhaustive methods reduce the computational complexity of exhaustive ones.
  Consequently, they allow for larger GWAS analysis at the cost of the
  possibility of not finding the target variant combination.
\end{itemize}

Prior to this work, the performance of exhaustive and non-exhaustive methods has
been studied thoroughly in~\cite{ponte-fernandez_evaluation_2020}. The paper
concluded that exhaustive methods are the only ones capable of identifying
epistasis interactions in the absence of marginal effects. Marginal effects
refer to the association effect that subgroups of the complete epistasis
interaction display with the trait under study. If associated variants do not
display marginal effects, non-exhaustive methods are ineffective and the only
known alternative is to exhaustively search the combination space. In spite of
that, due to implementation constraints, the majority of the proposed exhaustive
methods limit the size of the epistasis interactions.

This work presents Fiuncho, an exhaustive epistasis detection tool that supports
interactions of any given order, and exploits all levels of parallelism
available in a homogeneous CPU cluster to accelerate the computation and make
it more scalable with the size of the problem. To the best of our knowledge,
the proposed implementation is faster than any other state-of-the-art CPU
method.

The text is organized as follows: Section~\ref{sec:related} covers related works
and highlights the different trends in exhaustive epistasis detection.
Section~\ref{sec:background} describes the association algorithm used, and
Section~\ref{sec:method} details the parallel epistasis search implemented.
Section~\ref{sec:evaluation} includes the evaluation of Fiuncho. And lastly,
Section~\ref{sec:conclusions} presents the conclusions reached and highlights
some future lines of work.

\section{Related work}\label{sec:related}

There is abundant literature dedicated to epistasis detection methods. This work
focuses specifically on the exhaustive approach to epistasis detection because
it is the only one that obtains results in the absence of marginal effects in
the data.

All exhaustive methods follow the same principle: examining every combination of
variants available in the data, and locating the most associated ones with the
phenotype under study. As a consequence of that, all exhaustive methods present
a computational complexity of \(O(n^k \cdot O_{at})\), with \(n\) being the
number of variants in the search, \(k\) the order of epistasis explored and
\(O_{at}\) the computational complexity of each individual association test.
The expression assumes that the number of combinations without repetition,
\(\binom{n}{k}\), is equivalent to \(n^k\), since the epistasis order \(k\) is
smaller than \(n-k\). This rigidity in the method itself has led to the
development of proposals with more innovation in the different architectures
used to tackle the problem than in the algorithmic approach to it.

Initially, exhaustive methods did not target a computer architecture in
particular. They were written in languages such as Fortran, Java or C, and could
be used in any computer. This is the case of MDR~\cite{hahn_multifactor_2003},
one of the most recognized exhaustive epistasis detection methods in the
literature. MDR was written in Java, allows for epistasis interactions of any
given order and supports multithreaded execution, although the performance
achieved is not ideal in modern computers. Since then, improving performance has
become the focal point of the exhaustive methods.

Currently, implementations are more tailored to a particular computer
architecture in order to exploit all the resources offered to speed up the
search. MPI3SNP~\cite{ponte-fernandez_fast_2020} and
BitEpi~\cite{bayat_fast_2021} are two examples of exhaustive methods that use
CPUs, or clusters of CPUs, to perform the search. MPI3SNP implements a
3-locus epistasis search using MPI, in combination with multithreading, to speed
up the computation using multiple computing nodes. BitEpi, on the other hand,
uses an alternative representation of the genotype information in memory,
introducing a tradeoff between the complexity of the association test and the
use of a more memory-intensive approach to the computation. BitEpi implements a
2, 3 and 4-locus epistasis search that also uses multithreading to speed up the
search. Furthermore, for the \textit{x86\_64} CPU architecture,
there are some publications that discuss AVX vector implementations of the
epistasis search~\cite{campos_2020_heterogeneous, ponte-fernandez_simd_2021}.

Aside from CPUs, GPUs and FPGAs are two architectures that have gained some
interest from researchers in the field. GPUs are a great fit due to the high
degree of parallelism that they offer and the embarrassingly parallel nature of
the epistasis search. There are a multitude of methods that fall under this
category, with SNPInt-GPU~\cite{wienbrandt_2021_snpint} being one of the latest
examples. Furthermore, with the introduction of tensor cores in the most recent
GPU microarchitectures there has been an effort made to exploit these new
instructions in the epistasis detection problem~\cite{nobre_2020_exploring}.
FPGAs have also been employed, with methods that support exhaustive 2 and
3-locus epistasis detection~\cite{gonzalez_2015_parallelizing,
kassens_2015_high}, and more recently, epistasis interactions of any given
order~\cite{ribeiro_2021_hedacc}.

Lastly, some authors have embraced this diversity in architectures with methods
that support heterogeneous systems in order to complete the epistasis search.
This includes methods written in architecture-agnostic languages so that the
same implementation can be compiled for different
hardware~\cite{nobre_2021_fourth}, as well as methods that exploit computing
systems with different architectures simultaneously, and thus taking advantage
of the benefits of each separate architecture, such as CPUs with
iGPUs~\cite{campos_2020_heterogeneous}, CPUs with
GPUs~\cite{nobre_2020_accelerating} and GPUs with
FPGAs~\cite{wienbrandt_2019_1000}.

This paper presents Fiuncho, a method targeting CPU architectures that combines
explicit vector implementations for \textit{x86\_64} CPUs with
multithread and MPI multiprocess computing to exploit all resources offered by
a \textit{x86\_64} CPU cluster. Furthermore, a portable
implementation using standard C++ is also included to support other CPU
architectures. The exhaustive search implemented contemplates epistasis
interactions of any order which, to the best of our knowledge, makes it the only
CPU method, besides MDR, that does not limit the size of the interactions,
although Fiuncho is significantly faster.

\section{Background}\label{sec:background}

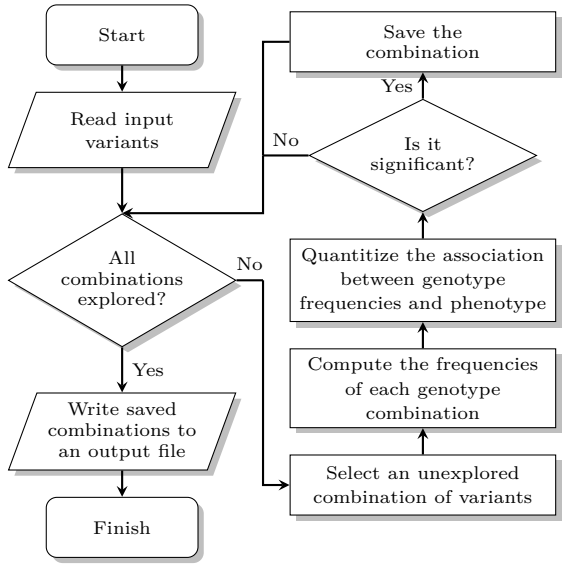
\begin{figure}
  \centering
  \begin{tikzpicture}[
        node distance=0.35cm,
        font=\footnotesize,
        startstop/.style={
                rectangle,
                fill=white,
                rounded corners,
                minimum width=2cm,
                minimum height=0.8cm,
                text centered,
                draw=black,
                drop shadow
            },
        io/.style={
                trapezium,
                fill=white,
                trapezium left angle=70,
                trapezium right angle=110,
                minimum width=3cm,
                minimum height=0.8cm,
                text centered,
                draw=black,
                drop shadow
            },
        process/.style={
                rectangle,
                fill=white,
                minimum width=3.5cm,
                minimum height=0.8cm,
                text centered,
                draw=black,
                drop shadow
            },
        decision/.style={
                diamond,
                fill=white,
                minimum width=3cm,
                minimum height=1.5cm,
                text badly centered,
                inner sep=-1ex,
                draw=black,
                drop shadow
            },
        arrow/.style={
                thick,
                ->,
                >=stealth
            }
    ]
    % Left column
    \node (start) [startstop] {Start};
    \node (in1) [io, below=of start.south] {\shortstack{
            Read input\\
            variants
        }};
    \node (dec1) [decision, below=of in1.south, yshift=-0.25cm] {
        \shortstack{
            All\\
            combinations\\
            explored?
        }};
    \node (out1) [io, below=of dec1.south, yshift=-0.25cm] {\shortstack{
        Write saved\\
        combinations to\\
        an output file
    }};
    \node (finish) [startstop, below=of out1.south] {Finish};
    % Right column
    % Coord to the right of the big decision node
    \coordinate [right=of dec1.east] (aux1);
    \node (pro1c) [process, right=of aux1] {
        \shortstack{
            Quantitize the association\\
            between genotype\\
            frequencies and phenotype
        }};
    \node (pro1b) [process, below=of pro1c] {
        \shortstack{
            Compute the frequencies\\
            of each genotype\\
            combination
        }};
    \node (pro1a) [process, below=of pro1b] {
        \shortstack{
            Select an unexplored\\
            combination of variants
        }};
    \node (dec2) [decision, above=of pro1c]
        {\shortstack{
            Is it\\
            significant?
        }};
    % Coord to the right of the second decision node
    \path let \p1=(dec2.west), \p2=(aux1)
        in coordinate (aux2) at (\x2,\y1);
    \node (pro3) [process, above=of dec2]
        {\shortstack{
            Save the\\
            combination
        }};

    % Direct arrows
    \draw [arrow] (start) -- (in1);
    \draw [arrow] (in1) -- (dec1);
    \draw [arrow] (dec1) -- node[anchor=west] {Yes} (out1.north);
    \draw [arrow] (pro1a) -- (pro1b);
    \draw [arrow] (pro1b) -- (pro1c);
    \draw [arrow] (pro1c) -- (dec2);
    \draw [arrow] (dec2) -- node[anchor=east] {Yes} (pro3);
    % Indirect arrows
    \draw [thick] (dec1) -- node[anchor=south] {No} (aux1);
    \draw [arrow] (aux1) |- (pro1a.west);
    \draw [thick] (dec2.west) -- node[anchor=south] {No} (aux2);
    \draw [thick] (pro3.west) -| (aux2);
    \draw [arrow] (aux2) |- (dec1.north);
    % \draw [thick] (pro3.west) -| (aux1);
    \draw [arrow] (out1.south) -- (finish);
\end{tikzpicture}
  \vspace{0.35cm}
  \caption{
      Flowchart of a typical exhaustive epistasis search.
  }\label{fig:exhaustive-flowchart}
\end{figure}

All exhaustive epistasis detection methods follow the same approach: enumerate
all combinations of variants for a particular order, test every combination for
association with the trait under study and report the relevant combinations.
Fig.~\ref{fig:exhaustive-flowchart} shows a flowchart of the process. Exhaustive
methods differ from one another in the association test used. Fiuncho, as
MPI3SNP~\cite{ponte-fernandez_fast_2020}, uses a Mutual Information (MI) based
association test. As can be seen in~\cite{ponte-fernandez_evaluation_2020}, MI
obtains a very good detection power.

This section briefly describes how the MI test operates, starting with the
construction of genotype tables to represent the genotype information of the
variants, followed by the computation of contingency tables to represent the
frequency of the genotype combinations corresponding to the selected variant
combination, and concluding with the MI test to assess the association between
the genotype frequencies and the phenotype.

\subsection{Constructing the genotype tables}

Genotype tables represent, in binary format, the genotype information of all
individuals under study for a particular variant or combination of variants.
They are a generalization of the binary representation introduced in
BOOST~\cite{wan_2010_boost} to simplify the computation of contingency tables
for second-order epistasis interactions. The tables contain as many columns as
individuals in the data, segregated into cases and controls, and as many rows as
genotype values a variant or combination of variants can show. Every individual
has a value of \texttt{1} in the row corresponding to its genotype and a
\texttt{0} in every other row. For a human population with biallelic markers,
each individual can have three different genotypes, and thus genotypes tables
contain \(3^k\) rows with \(k\) being the number of variants in combination
represented.

Genotype tables are not only used to represent the information of a variant, but
also to segment the individuals into different groups by their phenotype and
genotype values and to represent the information of multiple variants in
combination. This makes them extremely useful later when computing the
frequencies of each genotype value. The construction of a genotype table for a
combination of multiple variants implies:
\begin{enumerate}[label={\alph*)}]
  \item the combination of the different rows of the tables
        corresponding to the individual variants, and
  \item the computation of the intersection of each combination of rows (or
        genotype groups) via bitwise AND operations.
\end{enumerate}

Fig.~\ref{fig:genotype-table} gives an example of two genotype tables for two
variants \textit{a} and \textit{b} for sixteen individuals (eight cases and
controls), and the table resulting from the combination of these two variants.

\begin{figure}
  \centering
  \begin{tikzpicture}[
        title/.style={
                font=\bfseries\footnotesize
            },
        comment/.style={
            font=\itshape\footnotesize
        },
        table/.style={
                matrix of nodes,
                nodes={
                    rectangle,
                    draw=black,
                    minimum width=1.75cm,
                    text centered,
                    font=\footnotesize\ttfamily,
                    drop shadow
                },
                row sep=-\pgflinewidth,
                column sep=-\pgflinewidth,
                every even row/.style = {nodes={fill=gray!20}},
                every odd row/.style = {nodes={fill=white}},
            }
    ]
    % Genotype table A
    \matrix[table] (a)
    {
        00101101 & 10101001 \\
        10000000 & 01000010 \\
        01010010 & 00010100 \\
    };
    % Labels of table A
    \node[comment] (a_col1_label) [above=0cm of a-1-1] {Cases};
    \node[comment] (a_col2_label) [above=0cm of a-1-2] {Controls};
    \node[comment] (a_row1_label) [left=0cm of a-1-1] {\(a_1\)};
    \node[comment] (a_row2_label) [left=0cm of a-2-1] {\(a_2\)};
    \node[comment] (a_row3_label) [left=0cm of a-3-1] {\(a_3\)};
    % Title of table A
    \node[title] (a_title) [
        above=0.45cm of a_row1_label.north west, anchor=west
    ]{
        Genotype table of variant \textit{a}:
    };

    % Genotype table C
    \matrix[table, right=1.5cm of a.north east, anchor=north west] (c)
    {
        00000000 & 00000001 \\
        00000100 & 00100000 \\
        00101001 & 10001000 \\
        10000000 & 00000000 \\
        00000000 & 00000010 \\
        00000000 & 01000000 \\
        01010000 & 00010100 \\
        00000010 & 00000000 \\
        00000000 & 00000000 \\
    };
    % Labels of table C
    \node[comment] (c_col1_label) [above=0cm of c-1-1] {Cases};
    \node[comment] (c_col2_label) [above=0cm of c-1-2] {Controls};
    \node[comment] (c_row1_label) [left=0cm of c-1-1] {\(a_1 \cap b_1\)};
    \node[comment] (c_row2_label) [left=0cm of c-2-1] {\(a_1 \cap b_2\)};
    \node[comment] (c_row3_label) [left=0cm of c-3-1] {\(a_1 \cap b_3\)};
    \node[comment] (c_row4_label) [left=0cm of c-4-1] {\(a_2 \cap b_1\)};
    \node[comment] (c_row5_label) [left=0cm of c-5-1] {\(a_2 \cap b_2\)};
    \node[comment] (c_row6_label) [left=0cm of c-6-1] {\(a_2 \cap b_3\)};
    \node[comment] (c_row7_label) [left=0cm of c-7-1] {\(a_3 \cap b_1\)};
    \node[comment] (c_row8_label) [left=0cm of c-8-1] {\(a_3 \cap b_2\)};
    \node[comment] (c_row9_label) [left=0cm of c-9-1] {\(a_3 \cap b_3\)};
    % Title of table C
    \node[title] (c_title) [
        above=0.6cm of c_row1_label.north west, anchor=west, text width=3.5cm
    ]{
        Genotype table of variant \textit{a} {\texttimes} variant \textit{b}:
    };

    % Genotype table B
    \matrix[table, left=1.5cm of c.south west, anchor=south east] (b)
    {
        11010000 & 00010101 \\
        00000110 & 00100010 \\
        00101001 & 11001000 \\
    };
    % Labels of table B
    \node[comment] (b_col1_label) [above=0cm of b-1-1] {Cases};
    \node[comment] (b_col2_label) [above=0cm of b-1-2] {Controls};
    \node[comment] (b_row1_label) [left=0cm of b-1-1] {\(b_1\)};
    \node[comment] (b_row2_label) [left=0cm of b-2-1] {\(b_2\)};
    \node[comment] (b_row3_label) [left=0cm of b-3-1] {\(b_3\)};
    % Title of table B
    \node[title] (b_title) [
        above=0.45cm of b_row1_label.north west, anchor=west
    ]{
        Genotype table of variant \textit{b}:
    };
\end{tikzpicture}
  \vspace{0.35cm}
  \caption{
    Example of two genotype tables of two different variants, \(a\) and \(b\),
    for eight cases and controls, and the combined genotype table of the two
    variants.
  }\label{fig:genotype-table}
\end{figure}

\subsection{Computing the contingency tables}\label{sec:background:ctable}

A contingency table is a type of table that holds the frequency distribution of
a number of variables, that is, the genotype and phenotype distributions for
this domain of application. These frequencies can be directly obtained by
counting the number of individuals in each of the phenotype and genotype groups
created by the genotype table. This implies counting the number of bits set, an
operation commonly known as a population count. Fig~\ref{fig:contingency-table}
shows the contingency tables of the example genotype tables included in
Fig.~\ref{fig:genotype-table}.

\begin{figure}
  \centering
  \begin{tikzpicture}[
    title/.style={
            font=\bfseries\footnotesize
        },
    comment/.style={
        font=\itshape\footnotesize
    },
    table/.style={
            matrix of nodes,
            nodes={
                rectangle,
                draw=black,
                minimum width=1.2cm,
                text centered,
                font=\footnotesize\ttfamily,
                drop shadow
            },
            row sep=-\pgflinewidth,
            column sep=-\pgflinewidth,
            every even row/.style = {nodes={fill=gray!20}},
            every odd row/.style = {nodes={fill=white}},
        }
]
% Genotype table A
\matrix[table] (a)
{
    4 & 4 \\
    1 & 2 \\
    3 & 2 \\
};
% Labels of table A
\node[comment] (a_col1_label) [above=0cm of a-1-1] {Cases};
\node[comment] (a_col2_label) [above=0cm of a-1-2] {Controls};
\node[comment] (a_row1_label) [left=0cm of a-1-1] {\(a_1\)};
\node[comment] (a_row2_label) [left=0cm of a-2-1] {\(a_2\)};
\node[comment] (a_row3_label) [left=0cm of a-3-1] {\(a_3\)};
% Title of table A
\node[title] (a_title) [
    above=0.6cm of a_row1_label.north west, anchor=west, text width=2.5cm
]{Contingency table of variant \textit{a}:};

% Genotype table C
\matrix[table, right=1.5cm of a.north east, anchor=north west] (c)
{
    0 & 1 \\
    1 & 1 \\
    3 & 2 \\
    1 & 0 \\
    0 & 1 \\
    0 & 1 \\
    2 & 2 \\
    1 & 0 \\
    0 & 0 \\
};
% Labels of table C
\node[comment] (c_col1_label) [above=0cm of c-1-1] {Cases};
\node[comment] (c_col2_label) [above=0cm of c-1-2] {Controls};
\node[comment] (c_row1_label) [left=0cm of c-1-1] {\(a_1 \cap b_1\)};
\node[comment] (c_row2_label) [left=0cm of c-2-1] {\(a_1 \cap b_2\)};
\node[comment] (c_row3_label) [left=0cm of c-3-1] {\(a_1 \cap b_3\)};
\node[comment] (c_row4_label) [left=0cm of c-4-1] {\(a_2 \cap b_1\)};
\node[comment] (c_row5_label) [left=0cm of c-5-1] {\(a_2 \cap b_2\)};
\node[comment] (c_row6_label) [left=0cm of c-6-1] {\(a_2 \cap b_3\)};
\node[comment] (c_row7_label) [left=0cm of c-7-1] {\(a_3 \cap b_1\)};
\node[comment] (c_row8_label) [left=0cm of c-8-1] {\(a_3 \cap b_2\)};
\node[comment] (c_row9_label) [left=0cm of c-9-1] {\(a_3 \cap b_3\)};
% Title of table C
\node[title] (c_title) [
    above=0.6cm of c_row1_label.north west, anchor=west, text width=3.5cm
] {Contingency table of variant \textit{a} {\texttimes} variant \textit{b}:};

% Genotype table B
\matrix[table, left=1.5cm of c.south west, anchor=south east] (b)
{
    3 & 3 \\
    2 & 2 \\
    3 & 3 \\
};
% Labels of table B
\node[comment] (b_col1_label) [above=0cm of b-1-1] {Cases};
\node[comment] (b_col2_label) [above=0cm of b-1-2] {Controls};
\node[comment] (b_row1_label) [left=0cm of b-1-1] {\(b_1\)};
\node[comment] (b_row2_label) [left=0cm of b-2-1] {\(b_2\)};
\node[comment] (b_row3_label) [left=0cm of b-3-1] {\(b_3\)};
% Title of table B
\node[title] (b_title) [
    above=0.6cm of b_row1_label.north west, anchor=west, text width=2.5cm
] {Contingency table of variant \textit{b}:};
\end{tikzpicture}
  \vspace{0.35cm}
  \caption{
    Contingency table examples using the same variants as in
    Fig.~\ref{fig:genotype-table}.
  }\label{fig:contingency-table}
\end{figure}

\subsection{Mutual Information test}

Once the contingency table is calculated, the only step left to assess the
association between the genotype distribution and the phenotype affliction is
computing the MI of the table. Considering two random variables \(X\) and \(Y\)
representing the genotype and phenotype variability, respectively, the MI can be
obtained as:

\begin{equation}
  MI(X;Y) = H(X) + H(Y) - H(X,Y)
\end{equation}

where \(H(X)\) and \(H(Y)\) are the marginal entropies of the two variables, and
\(H(X,Y)\) is the joint entropy. Marginal entropies of one and two variables are
obtained as:

\begin{equation}
  H(X) = - \sum_{x \in X} p(x) \log p(x)
\end{equation}

\begin{equation}
  H(X,Y) = - \sum_{x,y} p(x, y) \log p(x, y)
\end{equation}

The computational complexity of constructing the genotype and contingency
tables, and applying the MI test is \(O(3^k \cdot m)\), with \(k\) being the
number of variants in combination tested, and \(m\) the number of individuals
represented in the tables.

\section{Parallel method}\label{sec:method}

Fiuncho implements a parallel exhaustive detection method using a static
distribution strategy. Given a collection of genotype variants from two groups
of samples (cases and controls), Fiuncho tests for association every combination
of variants for a particular interaction order using the association test
presented in Section~\ref{sec:background}, and reports the most associated
combinations. To do this, Fiuncho combines three different levels of
parallelism:

\begin{itemize}
  \item Task parallelism: the search method is divided into independent tasks
        that are distributed among the processing resources available in a
        cluster of CPUs. MPI multiprocessing and multithreading are used for
        the implementation.

  \item Data and bit-level parallelism: each task exploits the Vector Processing
        Units (VPUs) by using the Single Instruction Multiple Data (SIMD)
        algorithm proposed in~\cite{ponte-fernandez_simd_2021}, including the
        explicit vector implementations for the \textit{x86\_64}
        CPU architecture of the three stages of the association test presented
        in Section~\ref{sec:background}. Furthermore, this algorithm uses 64-bit
        word arrays to represent each of the rows of the genotype tables, and as
        a consequence of that, each intersection operation (bitwise AND) works
        with 64 samples at once.
\end{itemize}

This section discusses the method used to exploit the task parallelism. It
starts by describing the distribution strategy followed in order to divide and
distribute the workload among the computational resources available, and
concludes with an algorithm that implements the epistasis search using the
presented strategy.

\subsection{Distribution strategy}\label{sec:method:par-strat}

In the epistasis search, the workload is implicitly divided by the combinations
themselves, and the association tests can be carried out in parallel using a
pool of processing units. Each association test involves the same computations.
However, many of the combinations share sub-combinations with one another, and
as such, many repeating computations concerning the construction of the genotype
tables can be avoided attending to how the combinations are scheduled on the
different units. For instance, when searching for fourth-order epistasis, the
analysis of the combinations with variants (1,2,3,4), (1,2,3,5), (1,2,3,6),
etc.\@ requires the construction of the same genotype table corresponding to the
pair (1,2) and the triplet (1,2,3). Therefore, assigning all these combinations
to the same unit will allow reusing the genotype tables of (1,2) and
(1,2,3) for all fourth-order combinations that contain them.

Fiuncho implements a static distribution strategy in which the combinations of
any given order \(k\) are distributed among homogeneous processing units using
the combinations of size \(k-1\), following a round-robin distribution of the
combinations sorted by ascending numerical order. In other words, every
combination of size \(k-1\) is scheduled among units, and every unit computes
all combinations of \(k\) variants starting with the given \(k-1\) prefix. This
strategy finds a middle ground between a good workload balance among
processing units and avoiding overlaps in computations between them. By
distributing the workload using the \(k-1\) combination prefixes we guarantee
that every combination of size \(k\) reuses the genotype tables of its prefix of
size \(k-1\), but it introduces an overlap between units during the calculation
of the tables of the \(k-1\) prefix. Nonetheless, repeating these calculations
results in a negligible overhead due to the exponential growth of the
combinatorial procedure, as the experimental evaluation included in
Section~\ref{sec:evaluation} proves.

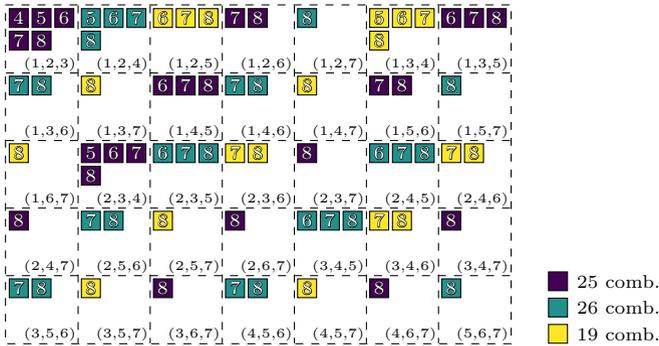
\begin{figure}
  \centering
  \begin{tikzpicture}[
        mat-pref/.style={
            matrix of nodes, nodes={
                rectangle,
                minimum width=0.95cm,
                minimum height=0.9cm,
                inner sep=0cm,
                outer sep=0cm,
                text height=0.80cm,
                text width=0.90cm,
                align=right,
                font=\tiny
            },
            inner sep=0cm,
            row sep=0cm,
            column sep=0cm
            },
        comb/.style={
                rectangle,
                draw=black,
                minimum width=0.25cm,
                minimum height=0.25cm,
                inner sep=0cm,
                outer sep=0cm,
                text=white,
                font=\footnotesize,
                text centered
            },
        g1/.style={
            fill=tikzcolor1
        },
        g2/.style={
            fill=tikzcolor2
        },
        g3/.style={
            fill=tikzcolor3
        },
        legend-label/.style={
            font=\footnotesize
        }
    ]
    % Define colors used in the style specification
    \definecolor{tikzcolor1}{rgb}{0.267,0.004,0.329};
    \definecolor{tikzcolor2}{rgb}{0.129,0.569,0.549};
    \definecolor{tikzcolor3}{rgb}{0.992,0.906,0.145};
    % Draw prefix matrix
    \matrix [mat-pref] (prefixes) {
        (1,2,3) & (1,2,4) & (1,2,5) & (1,2,6) & (1,2,7) & (1,3,4) & (1,3,5) \\
        (1,3,6) & (1,3,7) & (1,4,5) & (1,4,6) & (1,4,7) & (1,5,6) & (1,5,7) \\
        (1,6,7) & (2,3,4) & (2,3,5) & (2,3,6) & (2,3,7) & (2,4,5) & (2,4,6) \\
        (2,4,7) & (2,5,6) & (2,5,7) & (2,6,7) & (3,4,5) & (3,4,6) & (3,4,7) \\
        (3,5,6) & (3,5,7) & (3,6,7) & (4,5,6) & (4,5,7) & (4,6,7) & (5,6,7) \\
    };
    % Draw outer borders
    \draw[black,dashed]
        (prefixes.north west) -- (prefixes.north east) --
        (prefixes.south east) -- (prefixes.south west) --
        (prefixes.north west);
    % Draw column borders
    \draw[black,dashed]
        foreach \x in {1,...,6}
        {(prefixes-1-\x.north east) -- (prefixes-5-\x.south east)};
    % Draw row borders
    \draw[black,dashed]
        foreach \x in {1,...,4}
        {(prefixes-\x-1.south west) -- (prefixes-\x-7.south east)};

    % Declare command that prints combination squares
    \newcommand{\DrawCombinations}[5]{
        % cell, name prefix, color style, first digit, count
        \path
            let
                \p1=(#1.north west)
            in
                node (#2-1) [comb, #3, anchor=north west]
                at (\x1+0.05cm-.5\pgflinewidth,\y1-0.05cm+.5\pgflinewidth)
                {\contour{black}{#4}};
        \ifnum #5>1
            \path
                foreach \x in {2,...,#5}{
                    \ifnum \x=4
                        node (#2-\x) [
                            comb, #3, below=0.05cm of #2-1.south, anchor=north
                        ] {\contour{black}{\fpeval{\x+#4-1}}}
                    \else
                        node (#2-\x) [
                            comb, #3, right=0.05cm of #2-\fpeval{\x-1}.east
                        ] {\contour{black}{\fpeval{\x+#4-1}}}
                    \fi
                };
        \fi
    }

    % Draw combination squares starting with 1
    \DrawCombinations{prefixes-1-1}{c123}{g1}{4}{5}
    \DrawCombinations{prefixes-1-2}{c124}{g2}{5}{4}
    \DrawCombinations{prefixes-1-3}{c125}{g3}{6}{3}
    \DrawCombinations{prefixes-1-4}{c126}{g1}{7}{2}
    \DrawCombinations{prefixes-1-5}{c127}{g2}{8}{1}
    \DrawCombinations{prefixes-1-6}{c134}{g3}{5}{4}
    \DrawCombinations{prefixes-1-7}{c135}{g1}{6}{3}
    \DrawCombinations{prefixes-2-1}{c136}{g2}{7}{2}
    \DrawCombinations{prefixes-2-2}{c137}{g3}{8}{1}
    \DrawCombinations{prefixes-2-3}{c145}{g1}{6}{3}
    \DrawCombinations{prefixes-2-4}{c146}{g2}{7}{2}
    \DrawCombinations{prefixes-2-5}{c147}{g3}{8}{1}
    \DrawCombinations{prefixes-2-6}{c156}{g1}{7}{2}
    \DrawCombinations{prefixes-2-7}{c157}{g2}{8}{1}
    \DrawCombinations{prefixes-3-1}{c167}{g3}{8}{1}
    % Draw combination squares starting with 2
    \DrawCombinations{prefixes-3-2}{c234}{g1}{5}{4}
    \DrawCombinations{prefixes-3-3}{c235}{g2}{6}{3}
    \DrawCombinations{prefixes-3-4}{c236}{g3}{7}{2}
    \DrawCombinations{prefixes-3-5}{c237}{g1}{8}{1}
    \DrawCombinations{prefixes-3-6}{c245}{g2}{6}{3}
    \DrawCombinations{prefixes-3-7}{c246}{g3}{7}{2}
    \DrawCombinations{prefixes-4-1}{c247}{g1}{8}{1}
    \DrawCombinations{prefixes-4-2}{c256}{g2}{7}{2}
    \DrawCombinations{prefixes-4-3}{c257}{g3}{8}{1}
    \DrawCombinations{prefixes-4-4}{c267}{g1}{8}{1}
    % Draw combination squares starting with 3
    \DrawCombinations{prefixes-4-5}{c345}{g2}{6}{3}
    \DrawCombinations{prefixes-4-6}{c346}{g3}{7}{2}
    \DrawCombinations{prefixes-4-7}{c347}{g1}{8}{1}
    \DrawCombinations{prefixes-5-1}{c356}{g2}{7}{2}
    \DrawCombinations{prefixes-5-2}{c357}{g3}{8}{1}
    \DrawCombinations{prefixes-5-3}{c367}{g1}{8}{1}
    % Draw combination squares starting with 4
    \DrawCombinations{prefixes-5-4}{c456}{g2}{7}{2}
    \DrawCombinations{prefixes-5-5}{c457}{g3}{8}{1}
    \DrawCombinations{prefixes-5-6}{c467}{g1}{8}{1}
    % Draw combination squares starting with 4
    \DrawCombinations{prefixes-5-7}{c567}{g2}{8}{1}

    % Legend entries
    \path
        let
            \p1=(prefixes-5-7.south east)
        in
            node (legend-3)[
                comb, g3, anchor=south west,
                label={[legend-label]east:19 comb.}
            ] at (\x1+0.5cm,\y1) {};
    \node (legend-2) [
            comb, g2, above=0.1cm of legend-3.north, anchor=south,
            label={[legend-label]east:26 comb.}
    ] {};
    \node (legend-1) [
        comb, g1, above=0.1cm of legend-2.north, anchor=south,
        label={[legend-label]east:25 comb.}
    ] {};
\end{tikzpicture}
  \vspace{0.35cm}
  \caption{
    Example of the distribution strategy, arranging combinations of four
    variants among three processing units. Each prefix of three variants
    (represented as large squares with dotted lines) is assigned to a unit
    (shown as different colors) following a round-robin distribution, and that
    unit tests for association every combination of four variants starting with
    the prefix (represented as small colored squares).
  }\label{fig:distribution}
\end{figure}

Fig.~\ref{fig:distribution} exemplifies this strategy, showing the distribution
of the computations resulting from a fourth-order search (\(k=4\)) of eight
variants using three processing units. The figure uses squares with dotted lines
to represent all prefixes of \(k-1=3\) variants derived from combining the eight
inputs, displayed in sorted order from left to right and top to bottom. Each
prefix square includes one or more colored squared in its interior, representing
a combination of four variants to be tested for association, and the colors
indicate the unit which will carry out its test. Every combination under the
same prefix is assigned to the same unit, guaranteeing that the genotype table
of the prefix is computed only once, and every prefix is assigned to one of the
three units following a round-robin distribution. At the same time, there are
small overlaps between the computations corresponding to the different prefixes.
For example, the prefix (1,2,3) and (1,2,4) require constructing the same
genotype table for the combination (1,2), and since they were assigned to
different units, the table will be constructed more than once. This strategy
assigns twenty-five, twenty-six and nineteen combinations to the three
processing units, respectively. Although it does not create the most balanced
distribution possible, the strategy does not require synchronization or
communication between units, takes the reuse of genotype tables into account and
achieves very good results for a more realistic input size.

\subsection{Algorithmic implementation}

With the previous distribution strategy in mind, Algorithm~\ref{alg:search}
presents the pseudocode for the parallel epistasis detection method. It follows
the Single Program Multiple Data (SPMD) paradigm in which all computing units
execute the same function, while each unit analyzes a different set of
variant combinations. The implementation combines MPI multiprocessing with
multithreading to efficiently exploit the computational capabilities of CPU
clusters. Every MPI process reads the input variants and stores each one in a
genotype table, maintaining the individual variant information replicated in
each process. After that, each MPI process spawns a number of threads that
execute the function presented over a different set of variant combinations. The
input data is provided to the different threads through shared memory, making an
efficient use of the memory inside each node. This procedure allows the parallel
strategy to be abstracted from the topology of the cluster, so that the workload
is assigned to each core partaking in the computation regardless of its
location.

\begin{algorithm}[!ht]
  \caption{
    {\sc Parallel\_search}: Tests a subset of variant combinations
    for association with the trait under study.
  }\label{alg:search}
  \begin{algorithmic}[1]
    \renewcommand*\Call[2]{\textproc{#1}(#2)}
    \renewcommand{\algorithmicrequire}{\textbf{Input:}}
    \renewcommand{\algorithmicensure}{\textbf{Output:}}
    \Require \(A\) -- Array of genotype tables representing \(n\) input variants
    \Require \(L\) -- List of variant combinations to analyse
    \Require \(B\) -- Size of the block of operations
    \Ensure List of the \(S\) highest ranking \(k\)-combinations

    \Function{sorted\textunderscore insert}{\textit{list, ct,
        mi}}\label{alg:search:9}
        \If{size of \(\mathit{list} < S\) \AOr \(mi > list[0]\)}
            \State Find first \(i\) so that \(list[i] > mi\)
            \State Insert \(\{ct, mi\}\) before \(i\)
            \If{size of \(\mathit{list} > S\)}
            \State Remove \(list[0]\)
            \EndIf
        \EndIf
    \EndFunction\label{alg:search:10}
    \Statex
    \State \(sorted\_list \gets\) Empty list
    \State \(ct \gets \) Array of \(B\) contingency tables
    \State \(b \gets 0\)
    \For{\(\{i_1,\dots,i_{k-1}\}\) \AIN \(L\)}\label{alg:search:2}
        \State \(gt \gets A[i_1]\)\label{alg:search:3}
        \For{\(j \gets 2\) \TO \(k-1\)}
            \State \(gt \gets\) \Call{combine}{\(gt, A[i_j]\)}
        \EndFor\label{alg:search:4}
        \For{\(j \gets i_{k-1} + 1\) \ATO \(n\)}\label{alg:search:5}
            \If{\(b = B\)}\label{alg:search:7}
            \For{\(l \gets 0\) \ATO \(B\)}
                \State\(mi \gets\) \Call{mutual\textunderscore information}{\(
                ct[l]\)}
                \State \Call{sorted\textunderscore insert}{\(
                sorted\_list, ct[l], mi\)}
            \EndFor
            \State \(b \gets 0\)
            \EndIf\label{alg:search:8}
            \State\(ct[b] \gets\) \Call{combine\textunderscore
                and\textunderscore popcount}{\(gt, A[j]\)}\label{alg:search:6}
            \State \(b \gets b + 1\)
        \EndFor
    \EndFor
    \For{\(l \gets 0\) \ATO \(b\)}\label{alg:search:11}
        \State\(mi \gets\) \Call{mutual\textunderscore information}{\(ct[l]\)}
        \State \Call{sorted\textunderscore insert}{\(sorted\_list, ct, mi\)}
    \EndFor\label{alg:search:12}
    \State \Return \(sorted\_list\)\label{alg:search:13}
\end{algorithmic}
\end{algorithm}

The input arguments to the function are the array \(A\) of \(n\) genotype tables
representing the individual variants, the list of variant combinations \(L\) to
analyze and the size \(B\) of the blocks in which the integer and floating-point
vector operations will be segmented. The list of combinations \(L\) is provided
as an iterator that traverses through the combinations assigned to each core
without the need of storing the list in memory. In turn, it returns the list of
combinations of \(k\) variants with the highest MI values.

Integer and floating-point vector operations are part of the vector functions
implementing the association test as presented
in~\cite{ponte-fernandez_simd_2021}. The function \textsc{combine} implements
the construction of a genotype table from two previous input genotype tables,
and the function \textsc{combine\_and\_popcount}
combines in one function the construction of a genotype table with the
computation of the contingency table from the previous genotype table using a
population count function as explained in Section~\ref{sec:background:ctable}.
These two functions are implemented using boolean and integer vector arithmetic.
The function \textsc{mutual\_information} implements the MI test,
and uses floating point vector arithmetic.

In addition, \textit{x86\_64} processors are known to reduce the
clock frequency attending to three factors: the number of active cores, the
width of the VPU used and the type of operations used. For instance, the
specification document~\cite{intel_xeon_specification} of the Intel Xeon
processor (the processor used during the evaluation) defines different base
frequencies attending to the number of cores and width of the AVX operations
used. Furthermore, this processor reduces its turbo frequencies if vector
floating-point arithmetic is used. To mitigate the impact of this frequency
reduction, the SIMD algorithm, previously referenced, segments the operations
into blocks so that each block can operate at a different
frequency~\cite{ponte-fernandez_simd_2021}, and the same technique is applied to
Algorithm~\ref{alg:search}.

The algorithm primarily consists of a \textit{for} loop that traverses the list
of variant combinations provided to the function (Line~\ref{alg:search:2}). The
loop begins by computing the genotype table for each combination prefix
\(\{i_1,\dots,i_{k-1}\}\). This is done in a progressive manner, starting with
the table of the first variant of the prefix \(i_1\), and adding one extra
variant to the genotype table at a time using the function \textsc{combine},
until the whole prefix is included in the table
(Lines~\ref{alg:search:3}--Lines~\ref{alg:search:4}). Once this table is
computed, every combination of \(k\) variants starting with the given prefix,
\(\{i_1,\dots,i_{k-1},j\}\) with \(j\in[i_{k-1},n-1]\), is examined using a
\textit{for} loop (Line~\ref{alg:search:5}). On each iteration, the genotype
frequencies of the combination are obtained through the function
\textsc{combine\_and\_popcount}, using the genotype
table of the prefix and the table of the variant \(j\)
(Line~\ref{alg:search:6}). The frequencies are stored in an array \(ct\) of
contingency tables. Only when \(B\) contingency tables are available, the loop
enters an \textit{if} branch where the table array \(ct\) is processed
altogether using a \textit{for} loop
(Lines~\ref{alg:search:7}--\ref{alg:search:8}), effectively separating the
floating-point vector computations of the \textsc{mutual\textunderscore
information} function from the genotype table construction operations. On each
iteration, a contingency table is processed by computing the MI of the table,
and its result is stored in a list of \(S\) elements, sorted by its MI value
using the auxiliary function defined in
Lines~\ref{alg:search:9}--\ref{alg:search:10}.

When the outermost \textit{for} loop ends, the remaining contingency tables
stored in the array \(ct\) are processed
(Lines~\ref{alg:search:11}--\ref{alg:search:12}) and the algorithm returns the
sorted list of the top-ranking \(S\) combinations (Line~\ref{alg:search:13}).

The beginning and the end are the only two points in the program requiring
synchronization among threads and MPI processes. Once all threads of a process
terminate, the different lists of top-ranking combinations kept in the shared
memory of the process are joined into one, then sorted by their MI value and
truncated to \(S\) combinations. Analogously, once all MPI processes have
assembled their joint lists, the results are gathered into a single joint list
through the MPI collective \texttt{MPI\_Gatherv}. This list is
then sorted by MI and truncated to \(S\) combinations again. To conclude, the
program writes the final list to a file and exits.

\section{Evaluation}\label{sec:evaluation}

This evaluation examines the proposed parallel method in terms of the balance
achieved by the parallel distribution, the overhead introduced by the overlap in
computations among the different processing units, the parallel efficiency
achieved for an increasing number of processing units and a comparison with
state of the art exhaustive epistasis detection software.
Table~\ref{tbl:scayle} describes the characteristics of each node from the
SCAYLE cluster used throughout the evaluation.

\begin{table}
  \centering
  \caption{
    Hardware and software description of the SCAYLE cluster nodes from the
    \texttt{cascadelake} partition.
  }\label{tbl:scayle}
  \begin{tabular}{lr}
    \toprule
    \multicolumn{2}{c}{SCAYLE node (cascadelake partition)} \\
    \midrule
    CPUs      & 2x Intel 6240 (36 cores) @ \qty{2.6}{GHz}   \\
    Memory    & \qty{192}{GB}                               \\
    Network   & Infiniband HDR @ \qty{100}{Gbps}            \\
    GPUs      & NVIDIA V100                                 \\
    OS        & CentOS 7.7                                  \\
    Kernel    & 3.10                                        \\
    Compiler  & GCC 11.2                                    \\
    Libraries & glibc 2.34                                  \\
              & OpenMPI 4.1.1                               \\
    \bottomrule
\end{tabular}
\end{table}

\subsection{Parallel distribution balance}

The distribution strategy presented in Section~\ref{sec:method:par-strat} does
not assign the same exact number of combinations of \(k\) variants to test for
association to every computing unit. Instead, the strategy makes a compromise
between the balance in combinations assigned and the reuse of intermediate
results.

\begin{figure}[!t]
  \centering
  \includegraphics[width=\columnwidth]{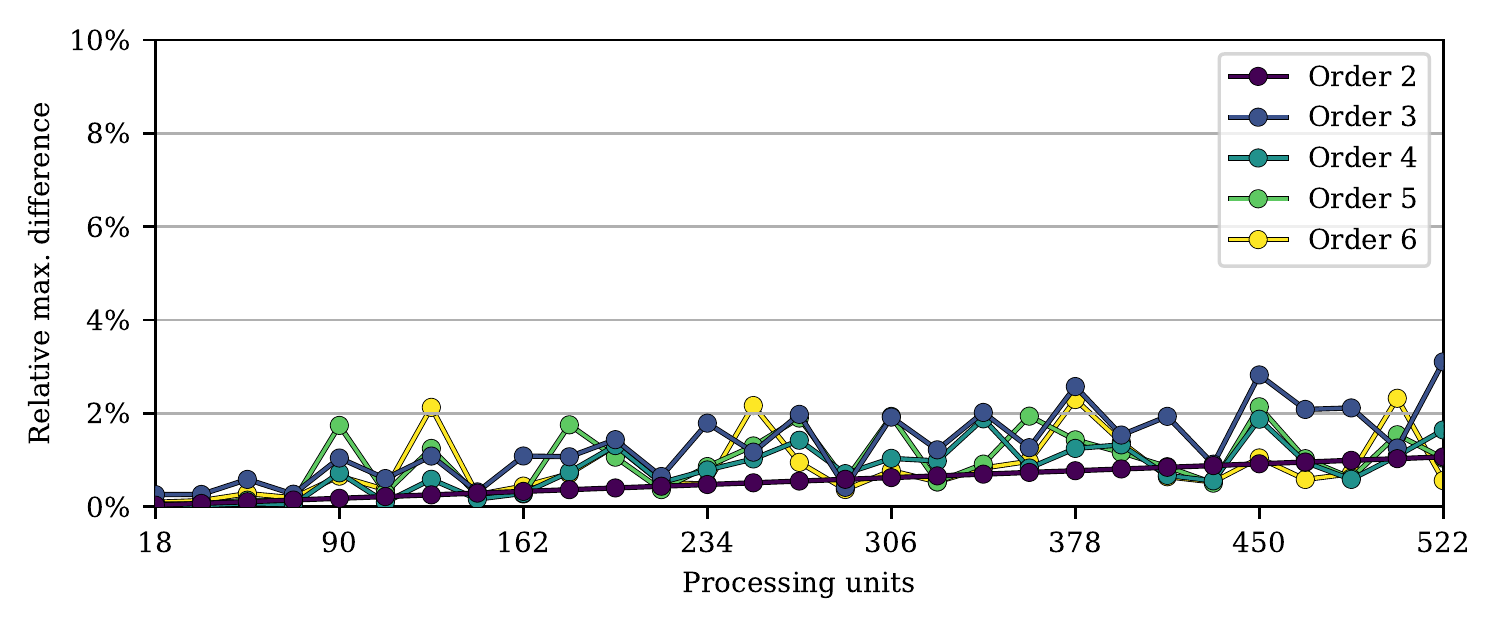}
  \caption{
    Maximum difference of assigned combinations to any processing unit from the
    average number of combinations assigned per unit, relative to the
    latter, for orders ranging from 2 to 6.
  }\label{fig:parallel_balance}
\end{figure}

In order to evaluate how good the designed strategy is,
Fig.~\ref{fig:parallel_balance} plots the maximum percentual difference between
the number of combinations assigned to a computing unit and the mean number of
combinations assigned to any unit, relative to the latter. It can be defined as:

\begin{equation}
  100 \: \frac {\max{d_i}-\binom{n}{k} / \mathit{p}}{\binom{n}{k} / \mathit{p}}
\end{equation}

with \(d_i\) being the number of combinations assigned to the unit \(i\), \(n\)
the number of variants, \(k\) the size of the combinations and \(\mathit{p}\)
the number of processing units used. The figure represents the differences in
workload distribution using combination sizes from 2 to 6 and a number of units
from 18 to 522. In order to keep a similar number of \(k\)-combinations, and
thus a similar distribution difficulty across combination orders, a number of
variants of \num{48828}, \num{1928}, 413, 172 and 100 were used for orders 2 to
6, respectively.

The results show that the proposed distribution keeps the differences under
\qty{3}{\percent} for every scenario tested. For scenarios with a larger variant
count, as is the case during the experimental evaluations of
Sections~\ref{sec:evaluation:speedup}~and~\ref{sec:evaluation:other}, the
differences in assigned workload are even smaller.

\subsection{Parallel overhead}

Although the distribution strategy takes into consideration the reuse of
genotype tables to avoid repeating the same computations in different processing
units, it certainly does repeat some operations during the construction of the
genotype table corresponding to the combination prefix assigned by the
distribution. In order to measure the overhead introduced, we compared the
elapsed time of a single-thread execution of the proposed implementation with an
alternative implementation of the same epistasis detection method that examines
every combination of variants sequentially and avoids repeating any calculation
pertaining to any genotype table.

Table~\ref{tbl:overhead} shows the overhead, measured as a percentage and
calculated as \(100 \cdot (T-T_{alt})/T_{alt}\), with \(T\) being the elapsed
time of the proposed implementation and \(T_{alt}\) being the elapsed time of
the alternative sequential implementation. The number of input variants selected
is inversely proportional to the order of the interaction in order to maintain
the runtime manageable, while the number of samples per variant was kept
constant (2048). The table omits the second and third-order overheads because,
for those combination sizes, the distribution strategy does not produce any
overlap in the computations associated with the calculation of genotype tables.
The results indicate that there is no significant difference between the two
elapsed times.

\begin{table}
  \centering
  \caption{
    Overhead of the parallel algorithm (run using a single CPU core) compared to
    a sequential implementation of the same operation, for interaction orders
    between four and six.
  }\label{tbl:overhead}
  \sisetup{
    table-format = 4.0,
    table-alignment = right,
    table-alignment-mode = format,
    table-text-alignment = right
}
\begin{tabular}{SSSS[table-format=4.2]S[table-format=4.2]S[table-format=-1.2]}
    \toprule
    {Order} & {Variants} & {Combinations} & {\(T\) (s)} & {\(T_{alt}\)(s)} & {Overhead (\%)} \\
    \midrule
    4       & 464        & 1906472876     & 1514.61     & 1526.48          & -0.78           \\
    5       & 152        & 632671880      & 1477.57     & 1453.21          & 1.68            \\
    6       & 76         & 218618940      & 1518.63     & 1506.17          & 0.83            \\
    \bottomrule
\end{tabular}

\end{table}

\subsection{Speedup and efficiency}\label{sec:evaluation:speedup}

This subsection evaluates the speedup and efficiency of Fiuncho using one and
multiple nodes. For both scenarios we selected a number of input variants
inversely proportional to the order of the interactions so that the elapsed
times of the analysis are similar, while the number of samples per variant was
kept constant at 2048.

\begin{figure}
  \centering
  \subcaptionbox{Observed speedups}{
    \includegraphics[width=.475\textwidth]{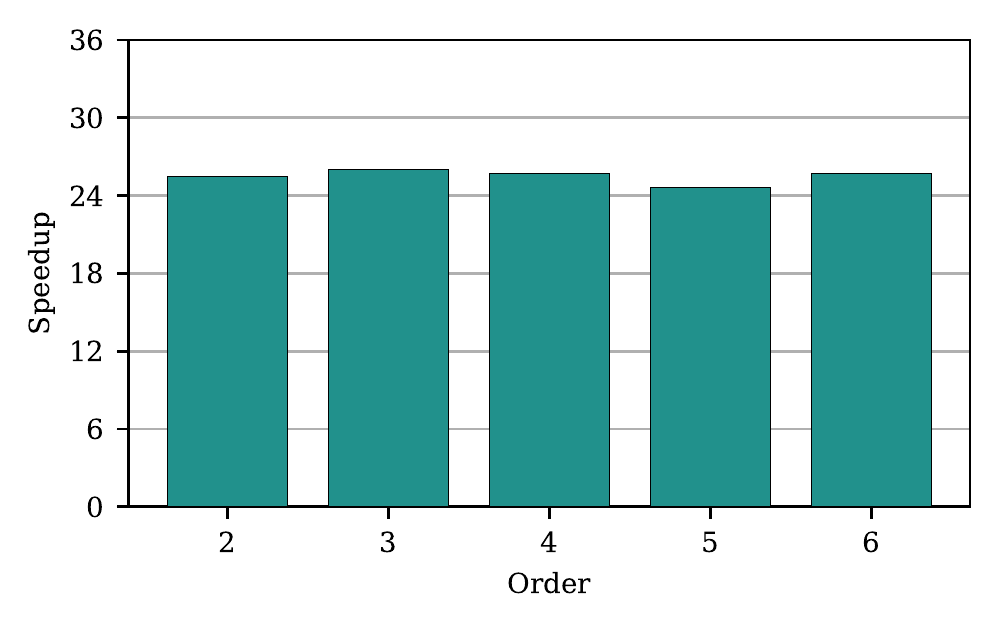}
  }
  \subcaptionbox{Freq.\ adjusted speedups}{
    \includegraphics[width=.475\textwidth]{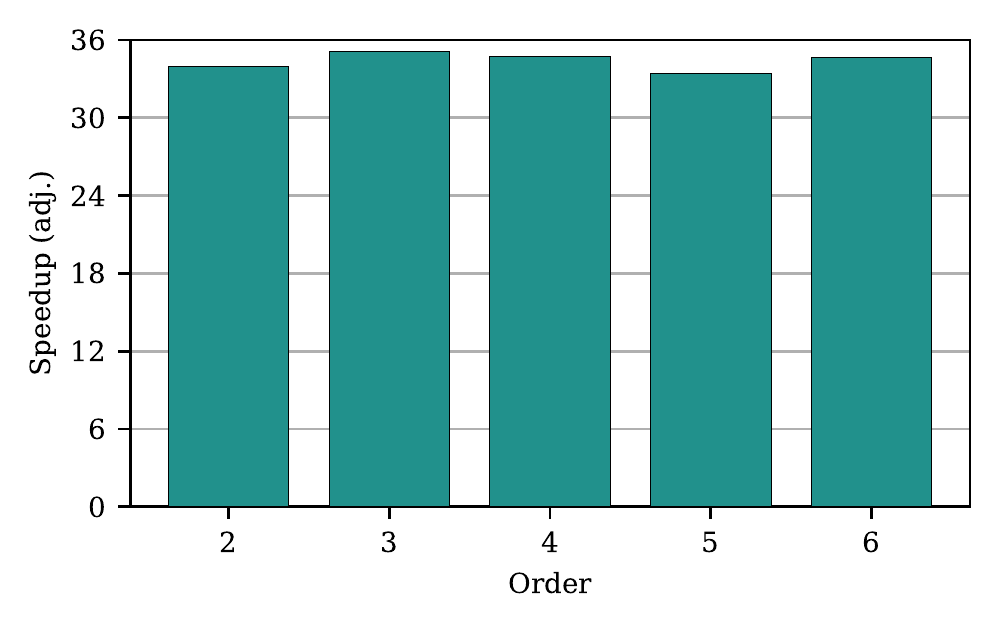}
  }
  \caption{
    Speedups of Fiuncho for multithread executions using 36 threads, compared
    to a single-thread run, representing both the observed and
    frequency-adjusted speedups.
  }\label{fig:speedups_intranode}
\end{figure}

Fig.~\ref{fig:speedups_intranode} represents the speedups obtained by Fiuncho
using a whole node (36 cores) when compared to single thread execution as seen
in Table~\ref{tbl:singlethread_intranode}, for epistasis orders ranging between
two and six. The figure shows two different metrics for the speedup: the
observed and the frequency-adjusted speedup. The observed speedup is calculated
as \(T_1/T_N\), with \(T_1\) being the elapsed time using a single CPU core and
\(T_N\) the elapsed time using \(N\) CPU cores. This metric is far from the
ideal efficiency, and this is due to the frequency scaling present in modern
processors. Intel CPUs, in particular, adjust their maximum core frequencies
attending to the number of active cores, with a larger frequency disparity if
AVX instructions are used~\cite{intel_xeon_specification}, as is the case with
Fiuncho. Therefore, to get a better grasp of the efficiency of the parallel
implementation, an adjusted speedup compensating for the discrepancy in average
CPU frequency is included in the figure, calculated as %
\(T_1/T_N \cdot F_1/F_N\), where \(F_1\) is the average single-core frequency
when Fiuncho uses a single core and \(F_N\) is the average multicore frequency
when \(N\) cores are used. The results for a single-node (36 CPU cores)
execution show very good efficiencies when the speedup is adjusted for the
frequency differences between single-core and multicore executions.

\begin{table}
  \centering
  \caption{
    Elapsed times of single-thread executions of Fiuncho for interaction orders
    between two and six.
  }\label{tbl:singlethread_intranode}
  \sisetup{
    table-format = 6.0,
    table-alignment = right,
    table-alignment-mode = format,
    table-text-alignment = right
}
\begin{tabular}{S[table-format=1.0]SS[table-format=11.0]S[table-format=4.2]}
    \toprule
    {Order} & {Variants} & {Combinations} & {Elapsed time (s)} \\
    \midrule
    2       & 184865     & 17087441680    & 2491.07            \\
    3       & 3246       & 5694987980     & 1612.47            \\
    4       & 464        & 1906472876     & 1514.61            \\
    5       & 152        & 632671880      & 1477.57            \\
    6       & 76         & 218618940      & 1518.63            \\
    \bottomrule
\end{tabular}
\end{table}

\begin{figure}
  \centering
  \subcaptionbox{2 nodes}{
    \includegraphics[width=.475\textwidth]{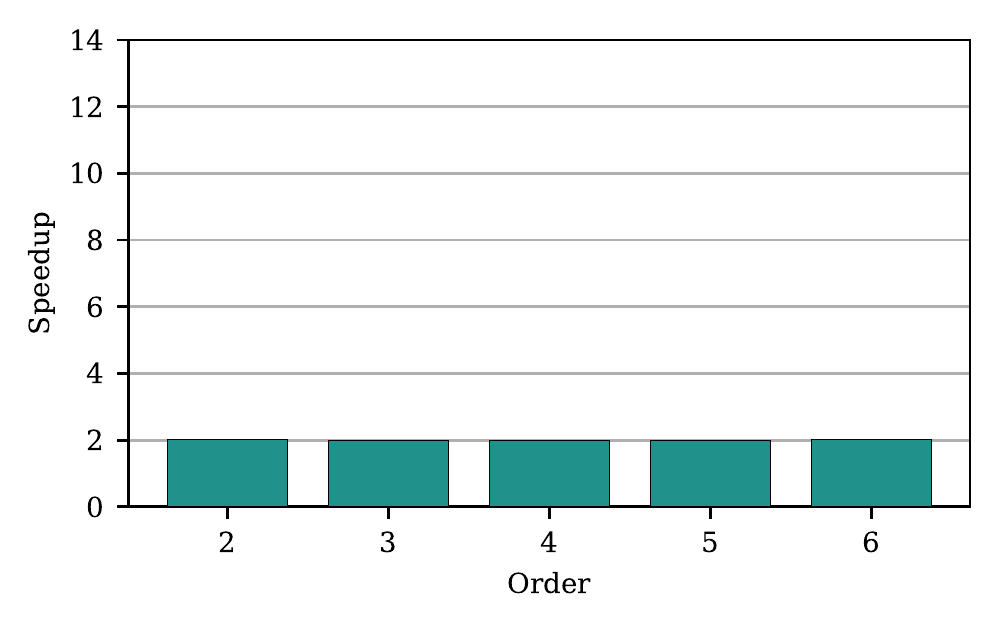}
  }
  \subcaptionbox{4 nodes}{
    \includegraphics[width=.475\textwidth]{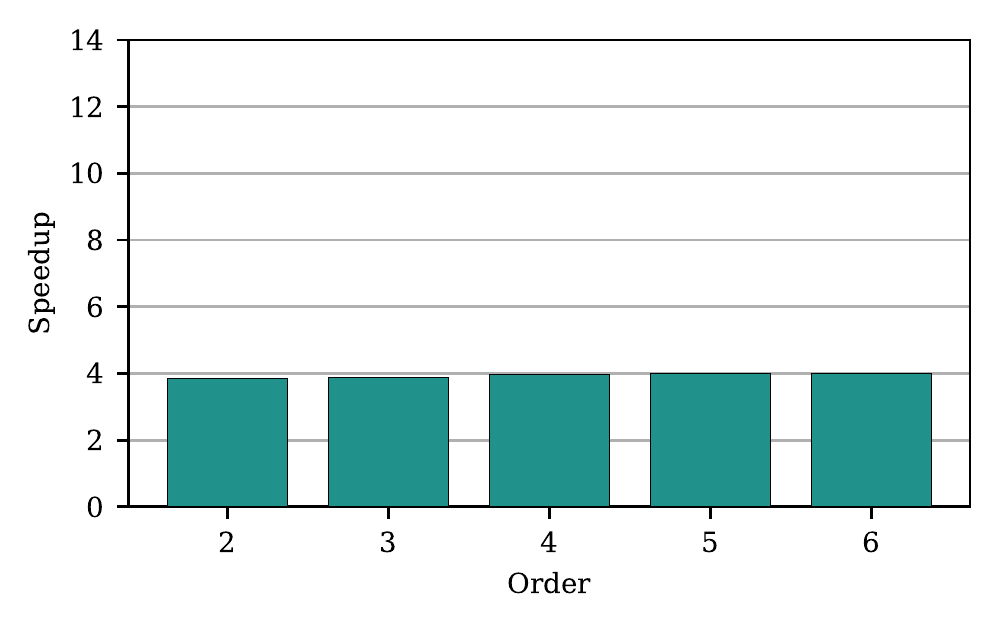}
  }
  \subcaptionbox{8 nodes}{
    \includegraphics[width=.475\textwidth]{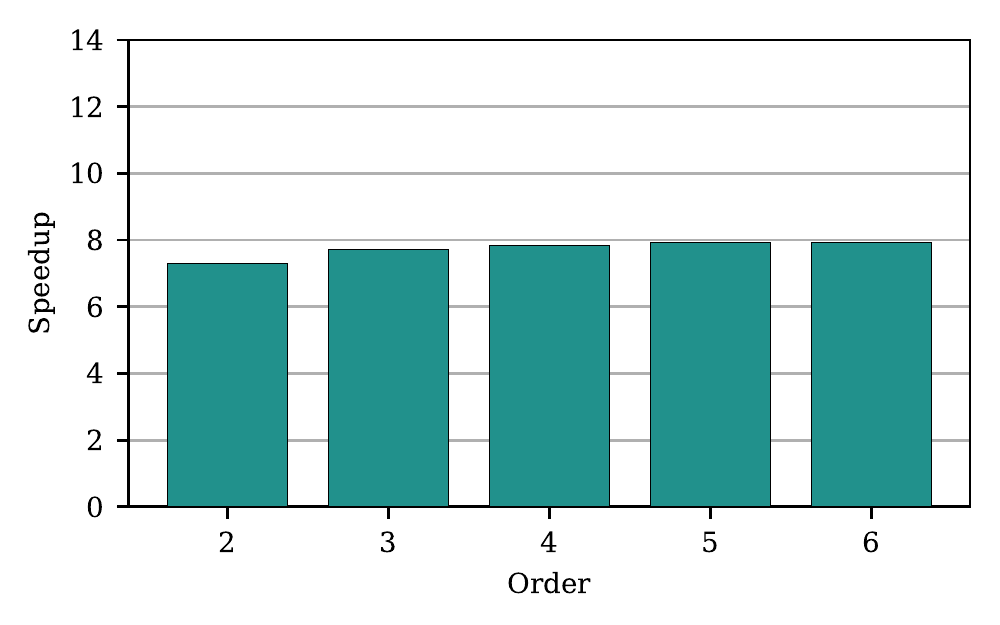}
  }
  \subcaptionbox{14 nodes}{
    \includegraphics[width=.475\textwidth]{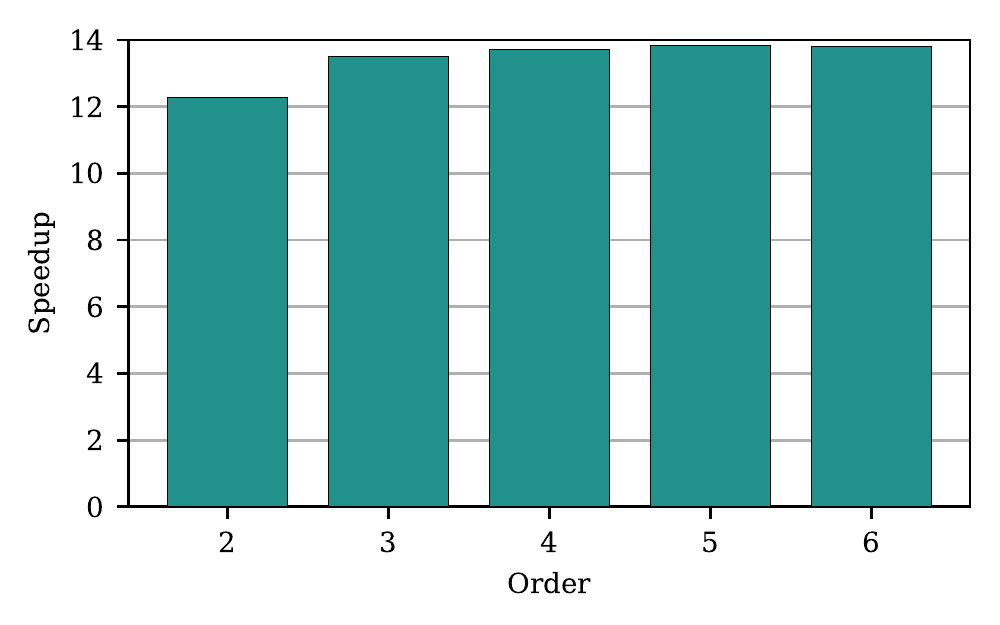}
  }
  \caption{
    Speedups of Fiuncho using 2, 4, 8 and 14 nodes with 36 threads per node,
    compared to a single-node execution.
  }\label{fig:speedups_internode}
\end{figure}

Fig.~\ref{fig:speedups_internode} shows the speedups obtained for multinode
executions using one MPI process per node with 36 threads each, comparing the
elapsed times obtained with a single-node run (36 cores) presented in
Table~\ref{tbl:singlenode_internode}. The datasets used in this second scenario
are substantially larger than those from Table~\ref{tbl:singlethread_intranode},
in order to keep the elapsed times over an hour long when 14 nodes (504 CPU
cores) are used. Here, in a multinode environment, there is no difference
between the average CPU frequency of the different nodes since all of them use
all the available CPU cores, and thus there is no need to include an adjusted
measure of the speedup. Again, results show very good efficiencies except for
the second-order interaction. This is due to the large input data for this
interaction order, sizing over \qty{29386}{MB} and read sequentially, thus
limiting the maximum speedup achievable.

\begin{table}
  \centering
  \caption{
    Elapsed times of single-node (36 cores) executions of Fiuncho for
    interaction orders between two and six.
  }\label{tbl:singlenode_internode}
  \sisetup{
    table-format = 7.0,
    table-alignment = right,
    table-alignment-mode = format,
    table-text-alignment = right
}
\begin{tabular}{S[table-format=1.0]SS[table-format=13.0]S[table-format=5.2]}
    \toprule
    {Order} & {Variants} & {Combinations} & {Elapsed time (s)} \\
    \midrule
    2       & 3755572    & 7052158645806  & 56261.70           \\
    3       & 28576      & 3888727096800  & 42539.30           \\
    4       & 2409       & 1399760565126  & 43176.00           \\
    5       & 561        & 454852770372   & 42594.70           \\
    6       & 223        & 159602946217   & 43103.40           \\
    \bottomrule
\end{tabular}
\end{table}

\subsection{Comparison with other software}\label{sec:evaluation:other}

Lastly, the performance of Fiuncho was compared with other exhaustive epistasis
detection tools from the literature: MPI3SNP~\cite{ponte-fernandez_fast_2020},
MDR~\cite{hahn_multifactor_2003} and BitEpi~\cite{bayat_fast_2021}. To do this,
we compared the elapsed times of all programs when looking for epistasis
interactions of orders two to four. In order to keep the elapsed time
constrained, multiple data sets were used containing a number of variants
inversely proportional to the order of the epistasis search and the hardware
resources used. The number of samples per variant, however, is \num{2048} for
all data sets. Since MDR is considerably slower than the rest of the programs
included, smaller data sizes were used for its evaluation.

Table~\ref{tbl:mpi3snp_results} compares the elapsed times of Fiuncho and
MPI3SNP, the tool previously developed by us. This program is limited to
third-order searches, thus the evaluation only considers this interaction order.
It implements MPI multiprocessing, so different scenarios were considered which
include single-thread, single-node and multinode configurations. Both MPI3SNP
and Fiuncho assign one MPI process per node, and create as many threads per
process as cores available in each node. The results show that Fiuncho is
significantly faster than MPI3SNP in all the evaluated scenarios.

\begin{table}
  \centering
  \caption{
    Elapsed time, in seconds, to complete an epistasis search both with MPI3SNP
    and Fiuncho, using a different number of nodes and CPU cores.
  }\label{tbl:mpi3snp_results}
  \sisetup{
    table-format = 5.0,
    table-alignment = right,
    table-alignment-mode = format,
    table-text-alignment = right
}
\begin{tabular}{SSS[table-format=12.0]SSS[table-format=4.2]S[table-format=5.2]}
    \toprule
    {Order} & {Variants} & {Combinations} & {Nodes} & {Cores} & {Fiuncho (s)} & {MPI3SNP (s)} \\
    \midrule
    3       & 3246       & 5694987980     & 1       & 1       & 1612.47       & 12868.42      \\
    3       & 8505       & 102498733260   & 1       & 18      & 2240.87       & 15312.84      \\
    3       & 10716      & 205033710860   & 1       & 36      & 2269.89       & 16046.58      \\
    3       & 13501      & 410062497750   & 2       & 72      & 2260.79       & 16085.50      \\
    3       & 17010      & 820134519120   & 4       & 144     & 2291.31       & 16186.12      \\
    \bottomrule
\end{tabular}
\end{table}

Table~\ref{tbl:bitepi_results} compares the results of BitEpi with Fiuncho.
BitEpi is a very novel program that only supports interaction orders between
two and four, thus the evaluation is restricted to those orders. Additionally,
BitEpi supports multithreading, therefore single-thread and multithread
scenarios are used. BitEpi uses a substantially different association test with
a time-complexity of \(O(n)\), while the association test used in Fiuncho has a
time-complexity of \(O(3^n)\). This can be observed in the results as a
shrinking difference between the elapsed times with the epistasis size. Despite
this, Fiuncho is still faster in all configurations tested. Furthermore, BitEpi
does not support multinode environments and can only exploit the hardware
resources of a single node, while Fiuncho can use as many resources as available
in order to reduce even further the elapsed time of the search.

\begin{table}
  \centering
  \caption{
    Elapsed time, in seconds, to complete an epistasis search both with BitEpi
    and Fiuncho, using different orders and number of CPU cores. The total
    workload between orders was kept as similar as possible.
  }\label{tbl:bitepi_results}
  \sisetup{
    table-format = 5.2,
    table-alignment = right,
    table-alignment-mode = format,
    table-text-alignment = right
}
\begin{tabular}{S[table-format=1.0]S[table-format=7.0]S[table-format=12.0]S[table-format=2.0]SS}
    \toprule
    {Order} & {Variants} & {Combinations} & {Cores} & {Fiuncho (s)} & {BitEpi (s)} \\
    \midrule
    2       & 184865     & 17087441680    & 1           & 2491.07       & 18090.91     \\
    2       & 784314     & 307573833141   & 18          & 3582.01       & 22294.39     \\
    2       & 1109187    & 615147345891   & 36          & 3797.39       & 23365.74     \\
    3       & 3246       & 5694987980     & 1           & 1612.47       & 5417.15      \\
    3       & 8505       & 102498733260   & 18          & 2240.87       & 7474.48      \\
    3       & 10716      & 205033710860   & 36          & 2269.89       & 7564.48      \\
    4       & 464        & 1906472876     & 1           & 1514.61       & 2202.65      \\
    4       & 954        & 34296318126    & 18          & 2101.60       & 3239.25      \\
    4       & 1134       & 68539472001    & 36          & 2111.74       & 3246.72      \\
    \bottomrule
\end{tabular}
\end{table}

Lastly, Table~\ref{tbl:mdr_results} compares the elapsed time of MDR with
Fiuncho, using a more limited number of variants than previous comparisons. MDR
is a relatively old program written in Java, but we decided to include it due to
its relevance in the field. It implements an epistasis search supporting
interactions of any order, although its elapsed time quickly becomes prohibiting
even with a reduced input size, so we decided to keep the interaction orders
between two and four. MDR supports multithreading, so single-thread and
multithread scenarios were considered in this evaluation. Results show a
massive difference in elapsed times, with an average speedup of 358 of Fiuncho
over MDR\@. This speedup could be increased even further if we considered
multinode scenarios for larger data sets, something that MDR does not support,
unlike Fiuncho.

\begin{table}
  \centering
  \caption{
    Elapsed time, in seconds, to complete an epistasis search both with MDR and
    Fiuncho, using different orders and number of CPU cores. The total workload
    between orders was kept as similar as possible.
  }\label{tbl:mdr_results}
  \sisetup{
    table-format = 5.2,
    table-alignment = right,
    table-alignment-mode = format,
    table-text-alignment = right
}
\begin{tabular}{SS[table-format=5.0]S[table-format=10.0]S[table-format=2.0]SS}
    \toprule
    {Order} & {Variants} & {Combinations} & {Cores} & {Fiuncho (s)} & {MDR (s)} \\
    \midrule
    2       & 9300       & 43240350       & 1       & 6.26          & 3571.77   \\
    2       & 39455      & 778328785      & 18      & 13.16         & 8656.46   \\
    2       & 55797      & 1556624706     & 36      & 16.04         & 10285.10  \\
    3       & 580        & 32350660       & 1       & 9.22          & 3204.88   \\
    3       & 1518       & 581842316      & 18      & 12.94         & 4710.82   \\
    3       & 1913       & 1164963436     & 36      & 13.34         & 6598.97   \\
    4       & 160        & 26294360       & 1       & 20.97         & 3767.28   \\
    4       & 328        & 473490550      & 18      & 29.04         & 4710.03   \\
    4       & 390        & 949173615      & 36      & 29.37         & 6491.59   \\
    \bottomrule
\end{tabular}
\end{table}

\section{Conclusions}\label{sec:conclusions}

This paper presents Fiuncho, an epistasis detection program written in C++, with
MPI directives and multithread support, that can be executed in CPU clusters. It
supports any interaction order, and implements an association testing method
based on the Mutual Information metric that has been proven to perform well in
epistasis detection~\cite{ponte-fernandez_evaluation_2020}. Fiuncho includes
explicit SIMD implementations of the association test calculations to exploit
the full computational capabilities of \textit{x86\_64}
processors.

Fiuncho exhibits exceptional performance, with a parallel strategy that balances
the workload remarkably well, obtaining computational efficiencies close to an
ideal growth with the hardware resources provided. When compared to existing
epistasis detection software, Fiuncho offers support for a wider scope of
application with no limit on the target epistasis size, and performs the fastest
of all programs considered in this study. For example, on average, Fiuncho is
seven times faster than its predecessor,
MPI3SNP~\cite{ponte-fernandez_fast_2020}, three times faster than
BitEpi~\cite{bayat_fast_2021} and 358 times faster than
MDR~\cite{hahn_multifactor_2003}\@. Moreover, the speedups over BitEpi and MDR
could be multiplied if larger experiments on multinode environments were
considered, as they are restricted to the hardware resources available in a
single node.

The main limitation of Fiuncho is its computational complexity, which makes its
cost prohibitive for large-scale studies and high interaction orders. For this
reason, future work should focus on improving the exhaustive strategy so that
its computational complexity can be reduced while not losing the epistasis
detection capabilities characteristic of these methods.

Fiuncho is distributed as open-source software, available to all the scientific
community in its Github
repository\footnote{\url{https://github.com/UDC-GAC/fiuncho}}.

\backmatter{}

\bmhead{Acknowledgments}

We would like to thank Supercomputación Castilla y León (SCAYLE), for providing
us access to their computing resources.

\section*{Declarations}

\begin{itemize}
\item Funding: this work was supported by the Ministry of Science and Innovation
      of Spain (PID2019{-}104184RB-I00 / AEI / 10.13039/501100011033), the Xunta
      de Galicia and FEDER funds of the EU (Centro de Investigación de Galicia
      accreditation 2019--2022, grant no. ED431G2019/01), Consolidation Program
      of Competitive Research (grant no. ED431C 2017/04), and the FPU Program of
      the Ministry of Education of Spain (grant no. FPU16/01333).
\item Conflict of interest/Competing interests: the authors have no competing
      interests to declare that are relevant to the content of this article.
\item Ethics approval: not applicable.
\item Consent to participate: not applicable.
\item Consent for publication: all authors have reviewed the study and consented
      to its publication.
\item Availability of data and materials: not applicable.
\item Code availability: the source code is distributed as open-source software,
      available in the repository \url{https://github.com/UDC-GAC/fiuncho}.
\item Authors' contributions: Conceptualization: all authors; Investigation:
      Christian Ponte-Fernández; Software: Christian Ponte-Fernández;
      Supervision: Jorge González-Domínguez and María J. Martín; Visualization:
      all authors; Writing -- original draft: Christian Ponte-Fernández; Writing
      -- review \& editing: all authors.
      % https://credit.niso.org/
\end{itemize}

\bibliography{references}

\end{document}